\documentclass[aps,prb,twocolumn,showpacs,superscriptaddress,floatfix]{revtex4-1}
\usepackage{amsfonts,hyperref}

\usepackage{graphicx}
\usepackage{amsmath}
\usepackage{amssymb}

\newcommand{\ket}[1]{\mbox{$| #1 \rangle$}}
\newcommand{\braket}[2]{\mbox{$\langle #1 | #2 \rangle$}}

\begin{document}
\title{Improved energy extrapolation with infinite projected entangled-pair states\\applied to the 2D Hubbard model}

\author{Philippe Corboz}
\affiliation{Institute for Theoretical Physics, University of Amsterdam,
Science Park 904, Postbus 94485, 1090 GL Amsterdam, The Netherlands}

\date{\today}

\begin{abstract}
An infinite projected entangled-pair state (iPEPS) is a variational tensor network ansatz for 2D wave functions in the thermodynamic limit where the accuracy can be systematically controlled by the bond dimension $D$. 
We show that for the doped Hubbard model in the strongly correlated regime (\mbox{$U/t=8$}, \mbox{$n=0.875$}) iPEPS yields lower variational energies than  state-of-the-art variational methods in the large 2D limit, which  demonstrates the competitiveness of the method. 
In order to obtain an accurate estimate of the energy in the exact infinite $D$ limit we introduce and test an extrapolation technique based on a truncation error computed in the iPEPS imaginary time evolution algorithm. The extrapolated  energies are compared with accurate quantum Monte Carlo results at half filling and with various other methods in the doped, strongly correlated regime.

\end{abstract}


\maketitle

\section{Introduction}
The accurate study of strongly correlated systems is one of the biggest challenges in condensed matter physics. A well known example is the 2D Hubbard model~\cite{Hubbard63} which potentially captures the relevant physics of the cuprate high-T$_c$ superconductors. Despite its simplicity and an enormous effort in trying to solve the model, the phase diagram of the Hubbard model is still controversial. Still, in recent years substantial progress has been achieved with a variety of different numerical methods (see e.g. Refs.~\onlinecite{white03,tocchio08,maier10, gull13,chen13,zheng15,deng15,leblanc15}), so that there is hope that the full solution of the Hubbard model may become within reach in the near future. For recent state-of-the-art benchmark results from various methods, see Ref.~\onlinecite{leblanc15}.

Solving systems in 1D is far better under control than in 2D, mostly thanks to the well-known density-matrix renormalization group (DMRG) method.~\cite{white1992}  DMRG has an underlying variational tensor network ansatz, called matrix product state (MPS), in which the wave function is efficiently represented by a trace over product of tensors. The accuracy of the ansatz can be systematically controlled by the bond dimension $D$ of the tensors, and one typically reaches extremely accurate results in 1D (and quasi 1D). DMRG can also be used to study 2D systems (typically on cylinders) by mapping the system onto a 1D problem with long-ranged interactions.~\cite{stoudenmire12} However, the computational cost scales exponentially with the width of the cylinder such that the approach is not scalable to large 2D systems.\footnote{Nevertheless, if finite size effects are not very large one can extract 2D physics in this way. For state-of-the-art examples, see e.g. Refs.~\onlinecite{Yan11,depenbrock12}.}

In order to overcome this exponential scaling 2D tensor network ans\"atze have been developed, such as projected entangled-pair states (PEPS,~\cite{verstraete2004,murg2007,Verstraete08,jordan2008} also called tensor product states~\cite{nishino01,nishio2004}) or the 2D multi-scale entanglement renormalization ansatz.~\cite{vidal2007-1,evenbly2009-2} These networks are designed in such a way that they  reproduce an area law scaling of the entanglement entropy which a large class of relevant ground states in 2D fulfill.\cite{eisert2010} The involved methods are technically more complicated than MPS-based approaches which is  one of the main reasons why it took several years to develop these methods. However,  recently  there have been substantial breakthroughs which clearly demonstrate the enormous potential of 2D tensor networks. For example, it was  shown for the $t$-$J$ model~\cite{corboz14_tJ} that infinite PEPS (iPEPS) - an ansatz for a state in the thermodynamic limit - yields lower variational energies than the state-of-the-art results from fixed-node Monte Carlo.~\cite{hu13} Another example is the Shastry-Sutherland model in a magnetic field,~\cite{corboz14_shastry} where iPEPS helped to gain a new understanding of the magnetization process, thanks to largely unbiased simulations.  

Thus, already  current (i)PEPS algorithms can outperform (or compete with)  the best variational methods for strongly correlated fermionic models like the $t$-$J$ model or also for frustrated spin systems (see e.g. Refs.~\onlinecite{corboz11-su4,wang11_j1j2,xie14}). 
 In this paper we show that the same is true also in the strongly correlated regime of the 2D Hubbard model, where we find lower variational energies than the best variational Monte Carlo results for large 2D systems.~\cite{leblanc15} 

One major difficulty in iPEPS simulations so far has been to obtain an accurate estimate of the energy in the exact, infinite $D$ limit. Typically the energy does not smoothly depend on the bond dimension $D$, making an extrapolation of the finite $D$ data to the infinite $D$ limit problematic. Accurate extrapolations become particularly important if several states at finite $D$ are strongly competing, as e.g. in the $t$-$J$ model,~\cite{corboz14_tJ} where uniform and stripe states exhibit almost the same energy at finite $D$. A precise estimate of the energy is crucial to identify the true ground state among these competing states. 

In this paper we propose and test an approach  to extrapolate the energy based on a truncation error $w$ which quantifies the degree of approximation  in the iPEPS imaginary time evolution algorithm. This quantity plays a similar role as the truncation error in conventional DMRG simulations which is typically used to extrapolate energies. Empirically we find here that the energy varies in a much smoother way with $w$ than with $1/D$ such that an extrapolation in $w\rightarrow 0$ 
yields an improved estimate of the exact ground state energy. 

We benchmark this extrapolation technique for the 2D Hubbard model, first in the exactly solvable $U/t=0$ limit (which is particularly challenging for iPEPS since the ground state is strongly entangled), then at finite $U/t$ at half-filling where we compare our results with accurate Quantum Monte Carlo results.\cite{leblanc15} Finally, we also provide an estimate of the energy in the doped, strongly correlated regime ($U/t=8$, $n=0.875$) and a comparison with various other methods from Ref.~\onlinecite{leblanc15}

This paper is organized as follows. In the next section we give  a short introduction to the iPEPS ansatz and the ground state algorithm based on  imaginary time evolution. In Sec. III we discuss ways to perform energy extrapolations with iPEPS, in particular, we explain how to compute a truncation error and use this quantity as an extrapolation parameter. 
In Sec. IV we present our  finite $D$ and extrapolated energies for the 2D Hubbard model and a comparison with other methods. Finally, in Sec. V we summarize our findings and give prospects for solving the Hubbard model. In appendix A we present additional results for the 2D Heisenberg model and a model of non-interacting spinful fermions with a pairing potential, to provide further evidence for the usefulness of the extrapolation technique based on the truncation error.

\section{iPEPS ansatz and method}
An infinite projected entangled-pair state (iPEPS)~\cite{verstraete2004,Verstraete08,jordan2008} is an efficient variational tensor network ansatz for two-dimensional states in the thermodynamic limit which obey an area law of the entanglement entropy (typically ground states of local Hamiltonians). The ansatz consists of a supercell of tensors which is periodically repeated on a lattice, with one tensor per lattice site.\cite{corboz2011} On the square lattice, each tensor has a physical index and four auxiliary indices which connect to the nearest-neighboring tensors.  The accuracy of the ansatz can be systematically controlled by the bond dimension $D$ of the auxiliary indices (i.e. each tensor contains $d D^4$ variational parameters where $d$ is the local dimension of a lattice site). An iPEPS with $D=1$ corresponds to a product state, and by increasing $D$ entanglement can be added in a systematic way.

For translational invariant states a supercell with only one single tensor can be used. If the translational symmetry is spontaneously broken, a supercell compatible with the symmetry breaking pattern is needed (e.g. for an antiferromagnetic state two different tensors for the two sublattices are required). Since in practice the structure of the ground state is not known in advance, we run simulations using different supercell sizes to check, which supercell yields the lowest variational energy. This approach also provides a way to determine several competing low-energy states, as for example done in the $t$-$J$ model,~\cite{corboz14_tJ} in which uniform and different stripe states have been found using different supercell sizes. In order to find the true ground state among these competing states, one needs to have an accurate estimate of the energy of each state in the infinite $D$ limit. However, a simple extrapolation in $D$ often fails to give an accurate estimate, due to the non-smooth dependence of the energy on $D$, and this is why it is important to find alternative ways to perform such extrapolations. Such an improved extrapolation technique will be presented in the next section.

The iPEPS wave function is evaluated by contracting the two-dimensional tensor network in a controlled approximate way.  In the present work we use a variant\cite{corboz14_tJ} of the corner-transfer matrix (CTM) method.~\cite{nishino1996, orus2009-1} The accuracy of the  contraction is controlled by the "boundary" dimension $\chi$, which we choose large enough (up to several hundreds) such that the resulting error is negligible (compared to the effect of the finite $D$). To increase the efficiency we make use of abelian symmetries.~\cite{singh2010,bauer2011} For an introduction to iPEPS, see e.g. Refs.~\onlinecite{corboz2010,phien15}. We note that 2D tensor networks have first been introduced for spin systems, and later extended to fermionic systems, see Refs.~\onlinecite{Corboz10_fmera, kraus2010, pineda2010,  barthel2009, shi2009, Corboz09_fmera,corboz2010,pizorn2010,gu2010}.

In order to obtain an approximate representation of the ground state for a given Hamiltonian $\hat H$, the tensors need to be \emph{optimized}, i.e. the best variational parameters have to be found. For iPEPS this is typically done by performing an imaginary time evolution of an initial (e.g. random) iPEPS. The evolution operator  is split into a product of two-site operators via a Trotter-Suzuki decomposition (assuming nearest-neighbor interactions),
\begin{equation}
\exp(-\beta \hat H) \approx \left( \prod_b \hat U_b \right)^n, \quad \hat U_b = \exp(-\tau \hat H_b),
\end{equation}
where the product goes over all nearest-neighbor bonds $b$ in the supercell, $\hat H_b$ is the Hamiltonian term on bond $b$, and  $\tau=\beta/n$ is a small imaginary time step. The imaginary time evolution is performed by sequentially multiplying the two-site operators $\hat U_b$ to the iPEPS and representing the resulting wave function again as an iPEPS, until convergence is reached. 
\footnote{In practice we use a second order decomposition, which is obtained by reverting the sequence of two-site operators at the even time steps.}

Let us consider the application of such a two-site operator $\hat U_b$ to two tensors $A$ and $B$ which are connected by the bond $b$. The resulting  state $|\Psi'_{A'B'}\rangle = \hat U_b |\Psi_{AB} \rangle$ can be represented by two new tensors $A'$ and $B'$ where the bond dimension on bond $b$ has increased from $D$ to $D' \leq d^2D$. 
For an efficient evolution the corresponding bond needs to be truncated back to the original bond dimension $D$, resulting in a truncated wave function $|\tilde \Psi_{\tilde A \tilde B} \rangle$. In the so-called full update~\cite{corboz2010} (or fast full update\cite{phien15}) this truncation is done by finding the new tensors $\tilde A$ and $\tilde B$ which  minimize the cost function 
\begin{equation}
\label{eq:C}
C = \min_{\tilde{A},\tilde{B}}|\, \mbox{\ket{\Psi'_{A'B'}}}-\mbox{\ket{\tilde \Psi_{\tilde{A}\tilde{B}}}}| =  \min_{\tilde{A},\tilde{B}}  \sqrt{d(\tilde{A},\tilde{B})}, \end{equation}
with
\begin{eqnarray}
d(\tilde{A},\tilde{B}) &=& \braket{\Psi'_{A'B'}}{\Psi'_{A'B'}}+\braket{\tilde\Psi_{\tilde{A}\tilde{B}}}{\tilde\Psi_{\tilde{A}\tilde{B}}}\\ \nonumber
&&-\braket{\Psi'_{A'B'}}{\tilde\Psi_{\tilde{A}\tilde{B}}}-\braket{\tilde\Psi_{\tilde{A}\tilde{B}}}{\Psi'_{A'B'}}.
\end{eqnarray}
Finding the new tensors can be solved in an iterative way, as explained e.g. in Refs.~\onlinecite{corboz2010,phien15}.\footnote{In practice only subparts of the tensors are updated to increase the efficiency.}

\section{Energy extrapolation with iPEPS}
Typically, for challenging problems, one does not reach convergence as a function of $D$ and one  needs to perform an extrapolation to the infinite $D$ limit to obtain an estimate of the true ground state energy. One possibility is to plot the energy as a function of $1/D$ and then trying to extrapolate the data to $1/D \rightarrow 0$. However, in practice the energy does not depend on $1/D$ in a smooth way which makes an accurate extrapolation difficult (see e.g. Refs.~\onlinecite{corboz14_shastry,Corboz13_shastry,Corboz12_su4,corboz11-su4} for examples).

Empirically one finds that the overall convergence of the energy goes faster than linear in $1/D$, such that a linear extrapolation in $1/D$ (using the largest few values of $D$) provides a lower bound $E_l$ of the true ground state energy. Since the method is variational, the energy for the largest value of $D$ corresponds to an upper bound~$E_u$. In practice a crude estimate of the energy can be obtained from the mean value $E_m=(E_u+E_l)/2$ with an error bar of $\Delta=(E_u-E_l)/2$. Examples of these estimates will be shown in the next section. While this approach can provide a reasonable guess of the exact energy, a more accurate and controlled extrapolation would be highly desirable.

In DMRG simulations energy extrapolations are typically much more accurate by extrapolating in the truncation error $\epsilon$, corresponding to the sum of the discarded squared singular values in the two-site variational optimization.\cite{schollwoeck2011} In simple words, the truncation error measures how far away the state is from the true ground state, which is reached if $\epsilon$ goes to zero. The question is now if a similar quantity could also be used in iPEPS simulations to improve energy extrapolations. The most natural way would be to implement a similar two-site variational optimization algorithm in iPEPS. However, the imaginary time evolution algorithm is more commonly used and more easy to implement, and we therefore aim to define a similar quantity within this approach.~\footnote{Another alternative would be to compute the variance of the Hamiltonian which is commonly used to perform extrapolations in other variational approaches. However, while computing the variance is in principle feasible by using projected entangled-pair operators, the computational cost would be rather large.}

Let us consider the cost function $C$ in Eq.~\eqref{eq:C}. For the true ground state, which is reached for sufficiently large $D$ and $\beta$, the cost function $C$ is zero.  However, if the true ground state is not reached because  $D$ is too small, then the cost function will reach a certain non-zero value $C(D, \beta\rightarrow \infty)$ for large $\beta$. For small $\tau$ the cost function depends linearly on the time step (to lowest order). We now define the quantity $w(D) = C(D, \beta\rightarrow \infty) / \tau$ which is independent of $\tau$ (for small $\tau$) and decreases monotonously with $D$. This quantity $w$ measures the truncation error when approximating an iPEPS with enlarged bond dimension $D'$ on a bond, $|\Psi'_{A'B'}\rangle$,  with a new iPEPS with smaller bond dimension $D$, $\ket{\tilde \Psi_{\tilde{A}\tilde{B}}}$. Thus, $w$ plays a similar role as the truncation error $\epsilon$ in DMRG simulations.

A priori we do not know how the energy depends on~$w$, in contrast to DMRG simulations where the energy converges linearly in the truncation error for sufficiently large bond dimensions (after a suitable number of finite-system sweeps). Nevertheless, we find here that the energy depends on $w$ in a much more regular way than on $1/D$, such that an extrapolation in $w$ using a polynomial fit to the data provides an improved estimate of the exact energy in the $w \rightarrow 0$ limit. This will be illustrated with several examples for the Hubbard model  in the next section.

\section{Benchmarks: 2D Hubbard model}
As a benchmark we consider the single-band 2D Hubbard model with only nearest-neighbor hoppings,
\begin{equation}\label{eq:hub}
\hat H=-t \sum_{\langle i,j, \sigma\rangle}  \left(\hat c_{i\sigma}^\dagger \hat c_{j\sigma}  + H.c.\right) +U\sum\limits_i \hat n_{i\uparrow}\hat n_{i\downarrow},
\end{equation}
where $\hat c^\dagger_{i\sigma}$ ($\hat c_{i\sigma}$) creates (annihilates) an electron with spin $\sigma=\left\{\uparrow, \downarrow\right\}$ on site $i$,  $\hat n_{i\sigma}=\hat c^\dagger_{i\sigma}\hat c_{i\sigma}$ is the number operator, $U$ is the on-site repulsion, and $t$ the hopping amplitude. 

In the following we first present the iPEPS results in the non-interacting case $U/t=0$ at half filling ($n=1$) which can be exactly solved. For the interacting case $U/t>0$ we make a comparison with several other methods from the recent state-of-the-art benchmark paper by LeBlanc et al.,\cite{leblanc15} including auxiliary-field quantum Monte Carlo (AFQMC), density-matrix embedding theory (DMET), DMRG, and fixed-node Monte Carlo (FNMC). 
We first consider the half filled case for $U/t=4$ and $U/t=8$ which can be accurately solved by AFQMC simulations since there is no sign problem at half filling. Finally, results for the doped case $n=0.875$ in the strongly correlated regime $U/t=8$ are presented and compared to the best available data. All energies are given in units of $t$.

\subsection{$U/t=0$ at half filling}
It is known that 2D free fermionic systems with a 1D Fermi surface have a multiplicative logarithmic correction to the area law of the entanglement entropy.~\cite{eisert2010} Since an iPEPS can only reproduce an area law this case poses a particular challenge and we do not expect to obtain the exact result for  finite $D$ (on an infinite lattice). Nevertheless, since the area law is only weakly violated (in contrast to a volume law), one can still obtain an approximation to the ground state and a variational estimate of the ground state energy. For example, for $D=16$ iPEPS yields an energy per site of \mbox{$E=-1.597$}. Compared to the exact result $-1.6211...$, this corresponds to a relative error of $\approx 1.5\%$.

\begin{figure}[]
\begin{center}
\includegraphics[width=1\columnwidth]{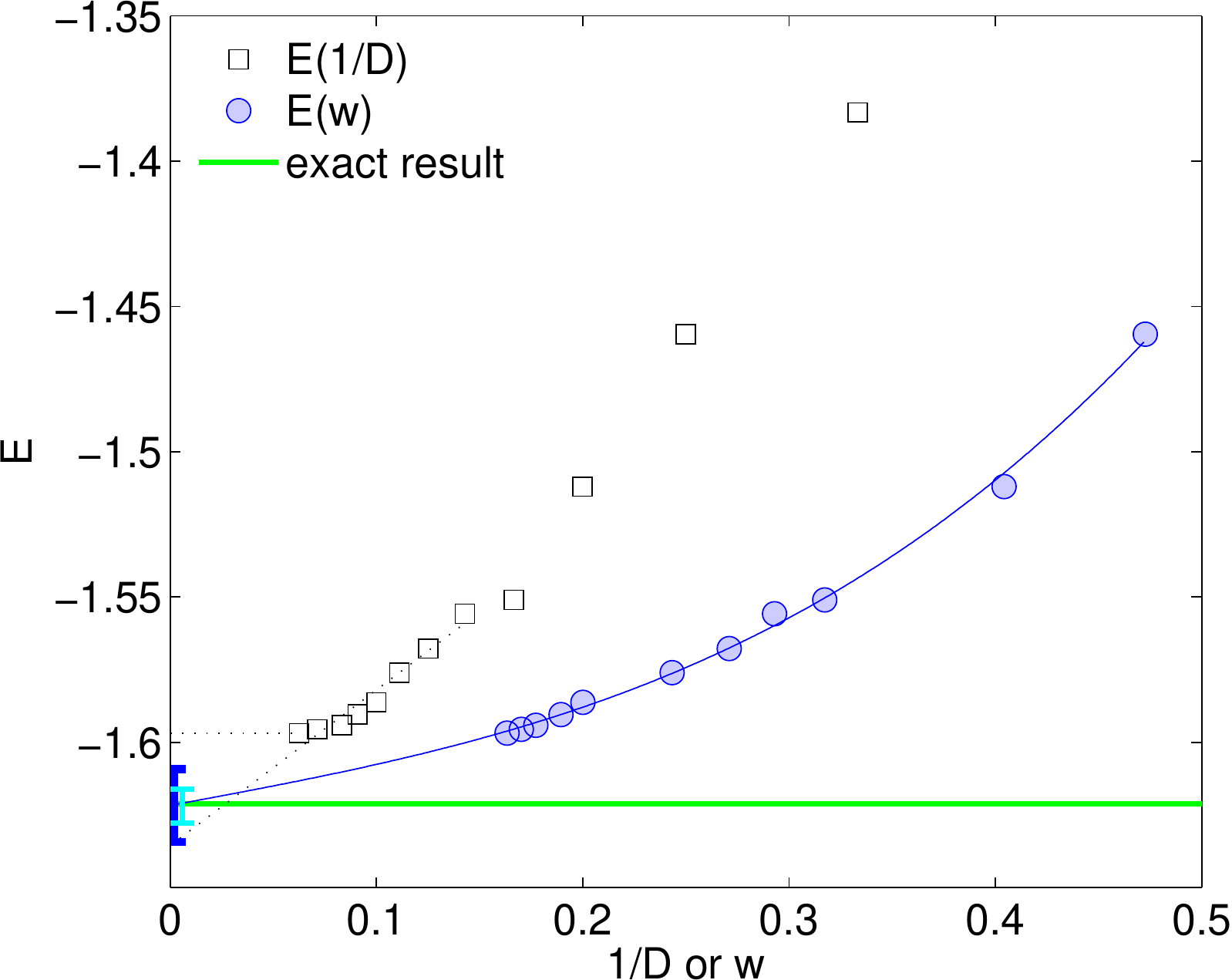} 
\caption{(Color online) iPEPS energies as a function of the inverse bond dimension (squares) and as a function of the truncation error $w$ (circles) for the non-interacting case ${U/t=0}$. }
\label{fig:U0}
\end{center}
\end{figure}

The iPEPS energies as a function of $1/D$ are shown in Fig.~\ref{fig:U0} (squares). One can clearly see how the variational energy improves upon increasing $D$. However, the dependence on $1/D$ is not very regular which makes an extrapolation in $1/D$ somewhat difficult. As discussed in the previous section we can obtain a lower bound,  
$E_l$, of the ground state energy by a linear extrapolation of the data in $1/D$, whereas the value at the largest $D$ corresponds to a variational upper bound, $E_u$. The range of energies estimated in this way is $[-1.597, -1.636]$, which includes the exact value. Based on this data we obtain an estimate $E_m=-1.616 \pm 0.019$. 

We next consider the iPEPS data plotted as a function of the average truncation error $w$ (circles in Fig.~\ref{fig:U0}). One can clearly observe a smoother dependence in $E(w)$ than in the $E(1/D)$ data. Fitting the data with a third-order polynomial  yields an energy $E_w=-1.6217$ in the limit $w\rightarrow 0$ which is close to the exact energy. As an estimate of the error we take  half the difference between lowest variational energy and the extrapolated value, $\Delta=(E_u  - E_w)/2 = 0.012$, shown by the dark blue error bar in Fig.~\ref{fig:U0}. As an alternative estimate we average over several fits using different ranges of data points, which yields $\tilde E_w = -1.6219$ with a standard deviation of $0.006$, shown by the light blue error bar in Fig.~\ref{fig:U0}.

Thus, even in the "worst-case" for iPEPS (i.e. free fermionic systems) we can obtain quite an accurate estimate of the ground state energy based on an extrapolation in the truncation error $w$.

\subsection{$U/t=4$ at half filling}
Next we consider the case $U/t=4$ at half filling, where the ground state is insulating and exhibits antiferromagnetic long-range order. Since there is no 1D Fermi surface as in the $U/t=0$ case, we expect the ground state to be less entangled (obeying an area law). This is reflected in a higher accuracy of the ground state energy obtained with iPEPS, and in a smaller truncation error $w$ compared to the $U/t=0$ case. For example, taking $D=16$ the relative error (compared to the extrapolated AFQMC result~\cite{leblanc15} $-0.8603 \pm 0.0002$) is of the order of $0.14\%$, i.e. an order of magnitude better than the $U/t=0$ result. The truncation error  is $w(D=16)\sim 0.05$ compared to $w(D=16)\sim 0.16$ in the $U/t=0$ case. 

\begin{figure}[]
\begin{center}
\includegraphics[width=1\columnwidth]{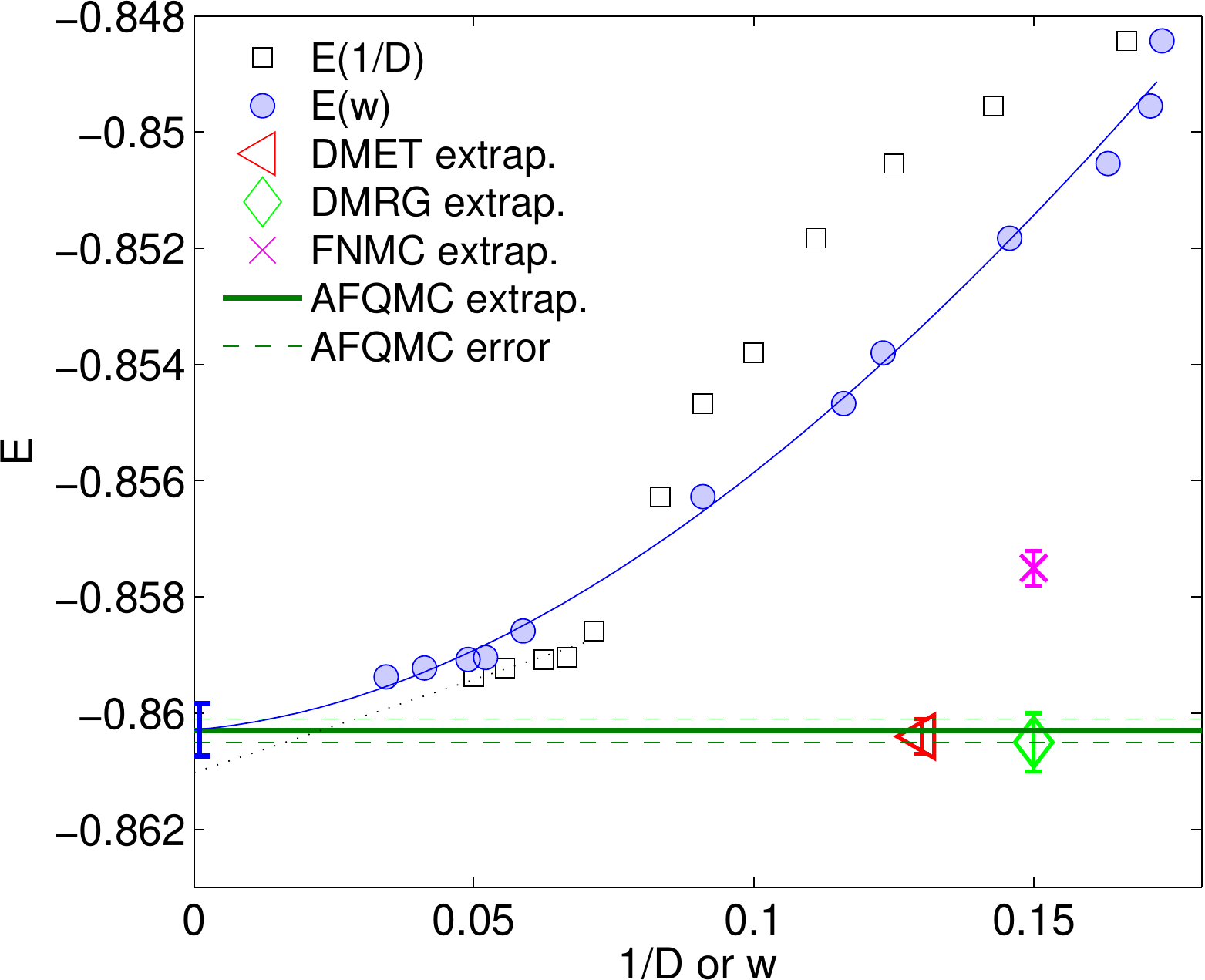} 
\caption{(Color online) iPEPS results for the energy for ${U/t=4}$ at half filling ($n=1$) as a function of $1/D$ (squares) and $w$ (circles), in comparison with the results from other methods (extrapolated to the thermodynamic limit). The reference value obtained from AFQMC is shown by the green line, with an  error bar indicated by the dashed lines. }
\label{fig:U4}
\end{center}
\end{figure}

The iPEPS energies exhibit a rather irregular behavior  as a function of $1/D$, shown by the squares in Fig.~\ref{fig:U4}. A linear fit using the five largest $D$ values yields the lower bound $E_l = -0.8610$. Computing an estimate based on the  $1/D$ extrapolation as in the $U/t=0$ case yields $-0.8602 \pm 0.0008$ in agreement with the AFQMC result.

The energies as a function of $w$ (circles)  exhibit a more regular behavior, as previously observed in the $U/t=0$ case. A third order polynomial fit including all data points yields $E_w=-0.8603 \pm 0.0005$. If we average again over several fits using different ranges of data points we obtain $\tilde E_w=-0.8604 \pm 0.0005$ in agreement with the AFQMC result. 

Our data is also in agreement with the extrapolated DMET and DMRG results, shown in Fig.~\ref{fig:U4}, with a comparable error bar.

\subsection{$U/t=8$ at half filling}
As we move away from the non-interacting limit, iPEPS becomes more accurate (in contrast to  weak-coupling approaches). 
For $U/t=8$, $D=16$ the energy is $-0.52415$. The relative error compared to the AFQMC result ($-0.5247\pm0.0002$) is small, only $0.1\%$. 
Thus, even without using any extrapolation we obtain already a remarkably accurate result in the thermodynamic limit.

The iPEPS estimate in the infinite $D$ limit based on the $1/D$ extrapolation is $-0.5246 \pm 0.0005$ (using the 4 largest $D$ values) and  $-0.5250 \pm 0.0008$ (using the 5 largest $D$ values), in agreement with the AFQMC result. 

\begin{figure}[]
\begin{center}
\includegraphics[width=1\columnwidth]{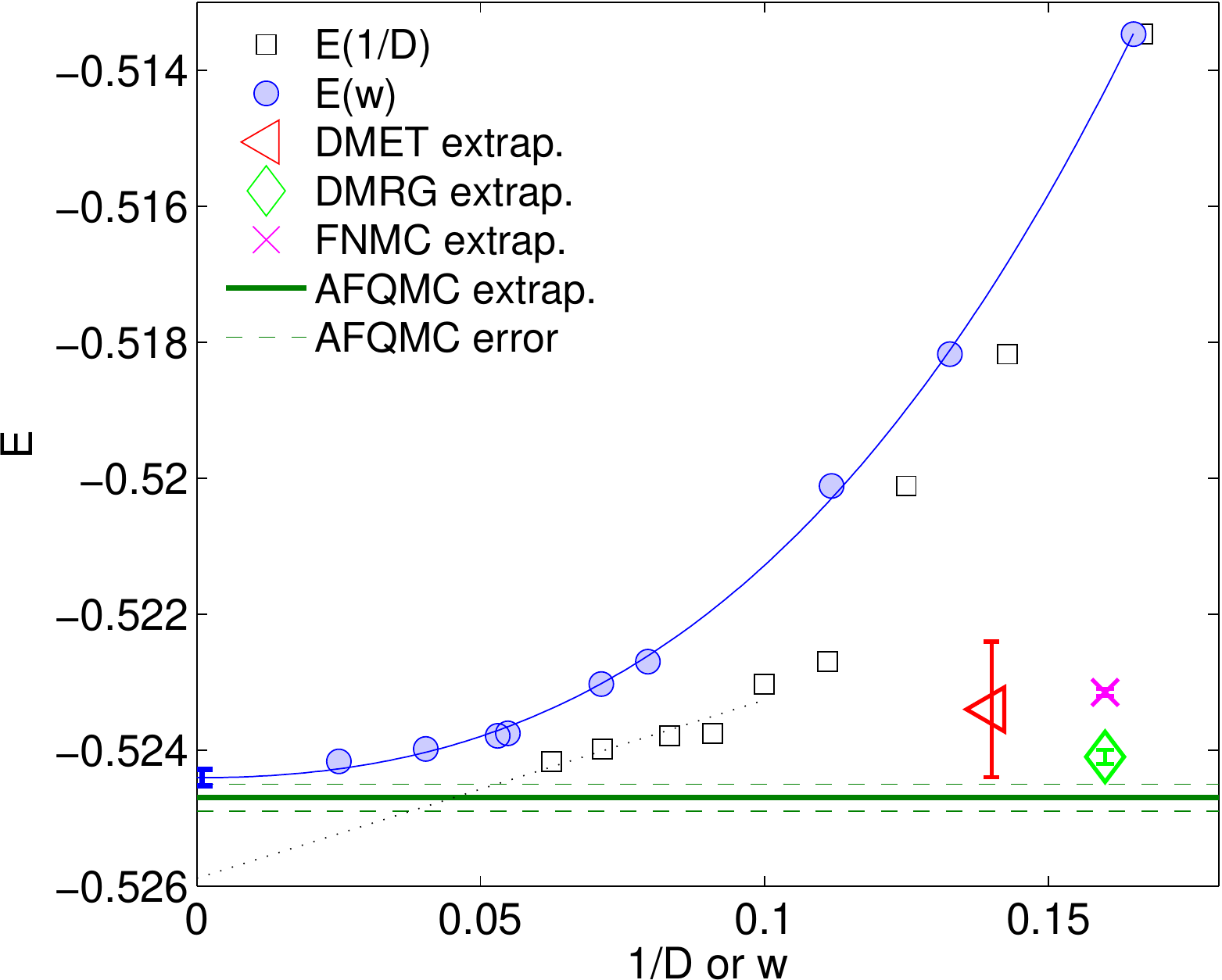} 
\caption{(Color online) iPEPS results for the energy for ${U/t=8}$ at half filling ($n=1$) in comparison with other methods.}
\label{fig:U8}
\end{center}
\end{figure}

Also here, the $E(w)$ curve is much smoother than the $E(1/D$) data, shown in Fig.~\ref{fig:U8}. A third-order polynomial fit yields \mbox{$E_w=-0.5244\pm0.0001$}, which is slightly higher, but still compatible with the AFQMC result. (A similar result is obtained by taking the average over several fits.)

Compared to the other methods iPEPS shows the best agreement with the AFQMC data: DMRG is slightly too high ($-0.5241\pm0.0001$), DMET has a large error bar ($-0.5234 \pm 0.001$) and the FNMC estimate is too high ($-0.52315 \pm 0.00005$).

\subsection{Doped case for $U/t=8$}
\begin{figure}[]
\begin{center}
\includegraphics[width=1\columnwidth]{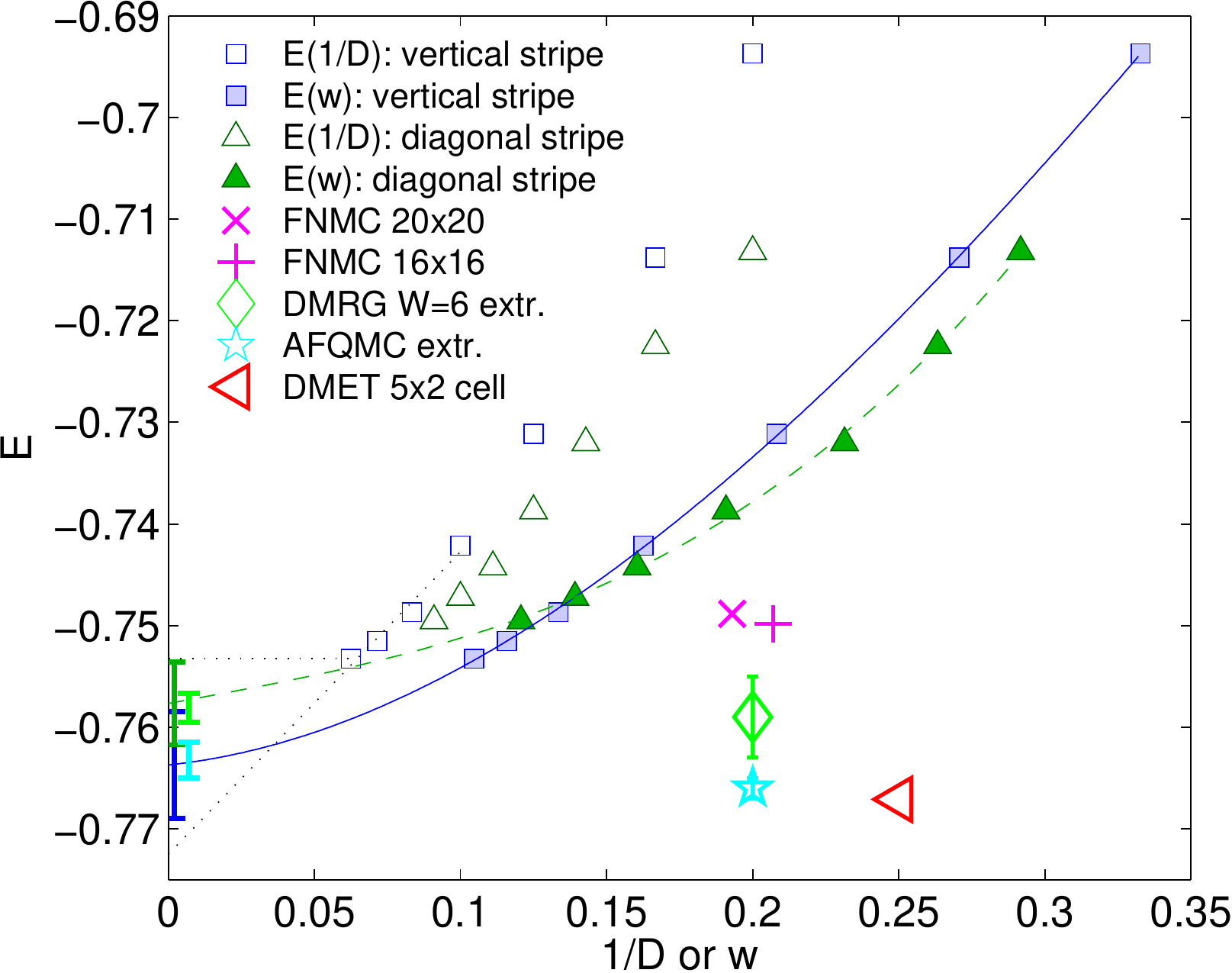} 
\caption{(Color online) iPEPS energy of a vertical stripe with period 5 and a diagonal stripe with period 16 for the doped Hubbard model in the strongly correlated regime ($U/t=8$, $n=0.875$) in comparison with other methods. }
\label{fig:U8d}
\end{center}
\end{figure}

Finally we present results in the doped, strongly correlated regime for $U/t=8$ and $n=0.875$ in Fig.~\ref{fig:U8d} for which AFQMC is no longer exact due to a strong sign problem. Determining the ground state in this regime is one of the fundamental open problems in the field of strongly correlated systems. The main reason why this is so challenging is that there are many  different states which are energetically very strongly competing (e.g. uniform superconducting states or different types of stripe states), which explains why different state-of-the-art methods yield different answers for the physics in this regime, see e.g. Ref.~\onlinecite{leblanc15}.

One of the main advantages of iPEPS is that the different competing states can be obtained by using different supercell sizes, and the properties of these states (e.g. order parameters) can then be studied individually for each low-energy state. This was done for example in Ref.~\onlinecite{corboz14_tJ} for the $t$-$J$ model (an effective model of the Hubbard model in the strongly interacting limit) where e.g. a uniform d-wave superconducting state with coexisting antiferromagnetic order is found by using a 2-site supercell, or period-5 stripe in a $5\times2$ supercell. 
In order to identify the true ground state among these competing states, one needs to compare their extrapolated energies, and this why an accurate extrapolation of the energy is important.

The systematic study of all possible supercell sizes and the discussion of the physics of all the competing states is beyond the scope of this paper and left for future work. Instead we illustrate here the advantage of the improved extrapolation technique by comparing the energies of  two different competing states: A vertical stripe state with a period 5 with charge and spin orders coexisting with superconductivity, and a (non-superconducting) diagonal stripe state with a period 16 with one hole per unit length per stripe. 

If we consider the energies plotted as a function of $1/D$, shown in Fig.~\ref{fig:U8d}, we can observe that for a given $D$ the diagonal stripe state has a lower variational energy than the vertical stripe state. However, since the energies of both states are still rapidly decreasing with increasing $D$ it is hard to predict which state is lower in energy in the large $D$ limit based on a $1/D$ extrapolation, which yields \mbox{$E_m = -0.763 \pm 0.010$} and \mbox{$E_m = -0.764 \pm 0.015$} for the vertical and diagonal stripe, respectively. These values lie very close and have a large overlapping error bar.

If, however, we use the truncation error $w$ as an extrapolation parameter, we can obtain a much clearer distinction between the two states. The extrapolation in $w$ yields \mbox{$E_w=-0.7637\pm0.005$} (\mbox{$E_w=-0.7577\pm0.004$}) for the vertical (diagonal) stripe; averaging over several fits using different ranges yields \mbox{$\tilde E_w=-0.7633 \pm 0.002$} (\mbox{$\tilde E_w=-0.7581 \pm 0.0014$}). Thus, in the large $D$ limit the vertical stripe state is favored over the diagonal stripe state. 

We next compare our results for the vertical stripe states with other methods from Ref.~\onlinecite{leblanc15}. Our best iPEPS variational energy for \mbox{$D=16$} is \mbox{$E=-0.75325$}. This is lower than the best variational result $-0.74884$ from FNMC for a $20\times 20$ system where the FNMC energies are increasing with system size. Thus, iPEPS clearly provides a lower variational energy in the thermodynamic limit than FNMC, which demonstrates the competitiveness of iPEPS in the doped, strongly correlated regime. 
%

The extrapolated iPEPS energy is comparable  to the result by (approximate) constrained path AFQMC ($-0.766 \pm 0.001$), the DMRG result (extrapolated in the truncation error) for a finite cylinder of width 6 ($-0.759\pm0.004$),  and the DMET result in a $5\times2$ cell ($-0.7671$).\footnote{G. K.-L. Chan, \textit{private communication}}

\section{Summary and prospects}
We presented iPEPS results for the energy of the 2D Hubbard model at half filling for $U/t=0$, 4, and 8, and away from half filling for $U/t=8$, $n=0.875$.  In the doped case the variational energies at large bond dimension are lower than the best available variational Monte Carlo results which  demonstrates that iPEPS is a very competitive variational method in the strongly correlated regime. It is in this challenging and physically relevant region where iPEPS (or 2D tensor networks in general) have the largest potential  to go substantially beyond the present state-of-the-art.

In order to obtain an estimate of the exact ground state energies we proposed to perform an extrapolation in the truncation error $w$, complementary to the (more crude) extrapolations in $1/D$, allowing us to compute ground state energies with a higher accuracy. At half filling the extrapolated results agree with the exact value in the non-interacting case, and with accurate AFQMC results for $U/t=4$ and $U/t=8$.
These extrapolations will play a key role to identify the true ground state among several competing states (e.g. stripe and uniform states\cite{corboz14_tJ}) which lie very close in energy. As an example we provided an estimate of the energy of a (vertical) period-5 stripe for $U/t=8$, $n=0.875$, and showed that the extrapolated energy is lower than the one of a period-16 diagonal stripe.

The present iPEPS data has been obtained using modest computational resources. We believe that with large scale parallel simulations using bond dimensions up to $D\sim20\dots24$, in combination with the present extrapolation technique, the ground state phase diagram in the strongly correlated regime ($U/t\ge6$) is accessible.
Combined with approaches which work best in the weakly correlated regime,~\cite{leblanc15,deng15} and with supporting results from other methods in the strongly correlated regime, the full solution of the 2D Hubbard model seems within reach. 

\acknowledgments
The author acknowledges useful discussions with J.~LeBlanc, E. Gull, S. R. White, and G. K.-L. Chan on the benchmark results for the Hubbard model.~\cite{leblanc15} 
This work is part of the D-ITP consortium, a program of the Netherlands Organization for Scientific Research (NWO) that is funded by the Dutch Ministry of Education, Culture and Science~(OCW).

\appendix
\section{Additional results}
In order to further demonstrate the usefulness of the energy extrapolation technique based on the truncation error $w$ we present additional results for the 2D Heisenberg model and for an exactly solvable model of non-interacting spinful fermions with a pairing potential in this appendix.

\subsection{2D Heisenberg model}
We consider the two-dimensional S=1/2 Heisenberg model on a square lattice with Hamiltonian,
\begin{equation}
\hat H=J \sum_{\langle i,j \rangle}  {\bf S_i S_j},
\end{equation}
where $\bf S_i$ is a spin-1/2 operator on site $i$. We set the coupling $J=1$. Since there is no negative sign problem this model can be  solved by Quantum Monte Carlo, and accurate estimates of the energy can by obtained by an extrapolation of the finite-size data to the thermodynamic limit. The Monte Carlo estimate is $E = -0.669437(5)$ from Ref.~\onlinecite{sandvik1997} obtained from linear system sizes up to $L=16$. A more precise estimate was presented in Ref.~\onlinecite{sandvik2010}, $E = -0.6694421(4)$, using larger system sizes. 

\begin{figure}[]
\begin{center}
\includegraphics[width=1\columnwidth]{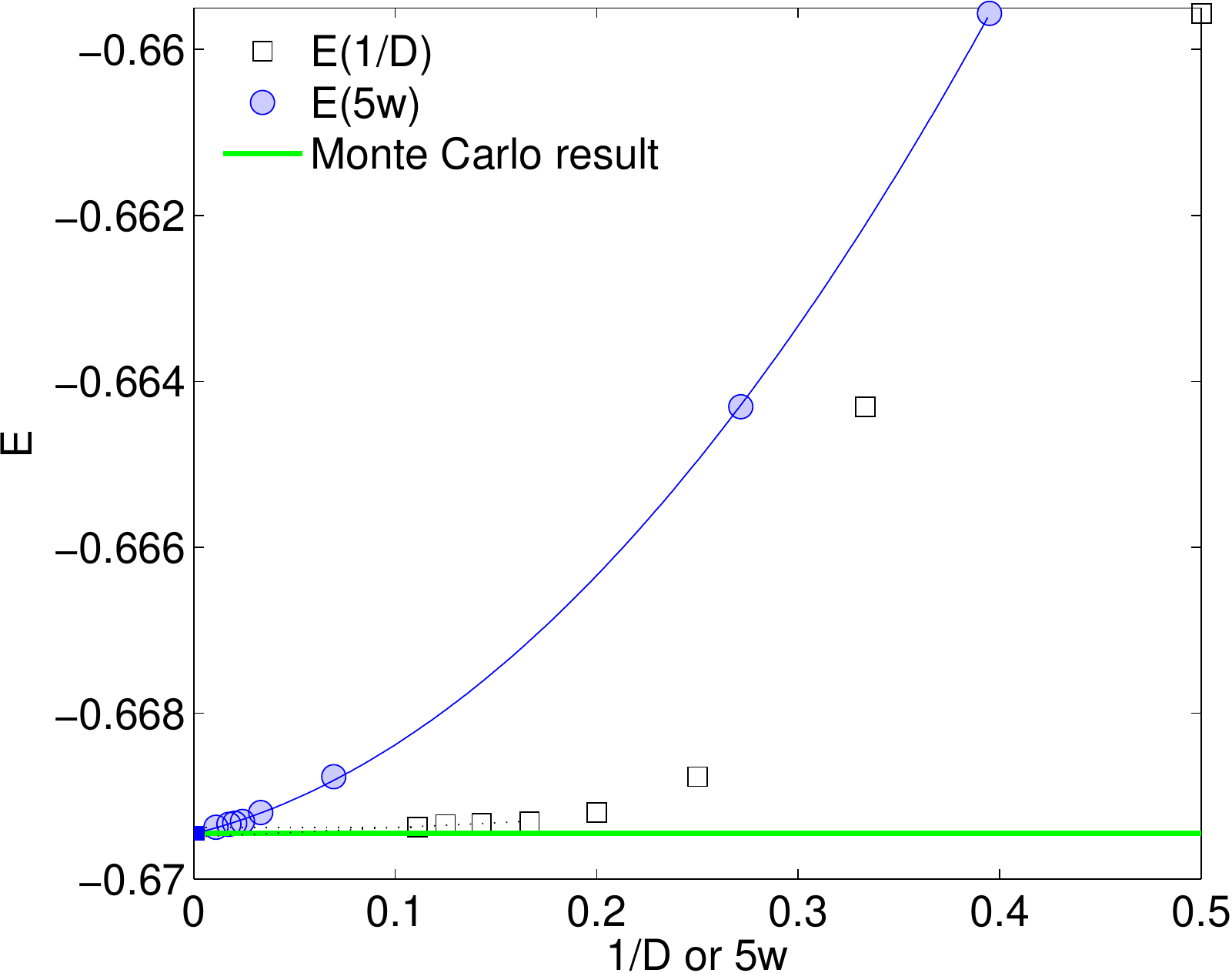} 
\caption{(Color online) iPEPS energy per site of the 2D Heisenberg model compared to the Quantum Monte Carlo result. Note that the truncation error on the x-axis has been rescaled by a factor 5 for better visibility.}
\label{fig:heis}
\end{center}
\end{figure}

Also in this example the iPEPS energies as a function of $1/D$ show a rather irregular behavior, as shown in Fig.~\ref{fig:heis}. Using an $1/D$ extrapolation yields $-0.66943(6)$ (using the four largest values of $D$), in agreement with the extrapolated QMC result.

The truncation errors in this example are considerably smaller than for the Hubbard model, indicating that the ground state is less entangled. Also in this case the dependence on $w$ is much smoother than on $1/D$. A second order polynomial fit including all data points yields $E_w = -0.66945(4)$. A similar result is obtained by averaging over several fits including different ranges of data points.

\subsection{Non-interacting spinful fermions with a pairing potential}
In this section we consider an exactly solvable model of spinful fermions with a nearest-neighbor hopping and a pairing potential term,
\begin{eqnarray}\label{eq:Hpair}
\hat H=&&-t \sum_{\langle i,j, \sigma\rangle}  \left(\hat c_{i\sigma}^\dagger \hat c_{j\sigma}  + H.c.\right) \nonumber
  \\  && + \sum_{\langle i,j \rangle} \gamma_{ij} \left( \hat c_{i\uparrow}^\dagger \hat c_{j\downarrow}^\dagger - \hat c_{i\downarrow}^\dagger   \hat c_{j\uparrow}^\dagger + H.c.  \right),
\end{eqnarray}
with $\gamma_{ij}$ the amplitude of the pairing potential. In the present example we set $t=1$, and $\gamma_{ij}=\pm 1$ for bonds oriented along the x- and y-direction, respectively. The exact energy in the thermodynamic limit is $-1.35494...$.

\begin{figure}[b]
\begin{center}
\includegraphics[width=1\columnwidth]{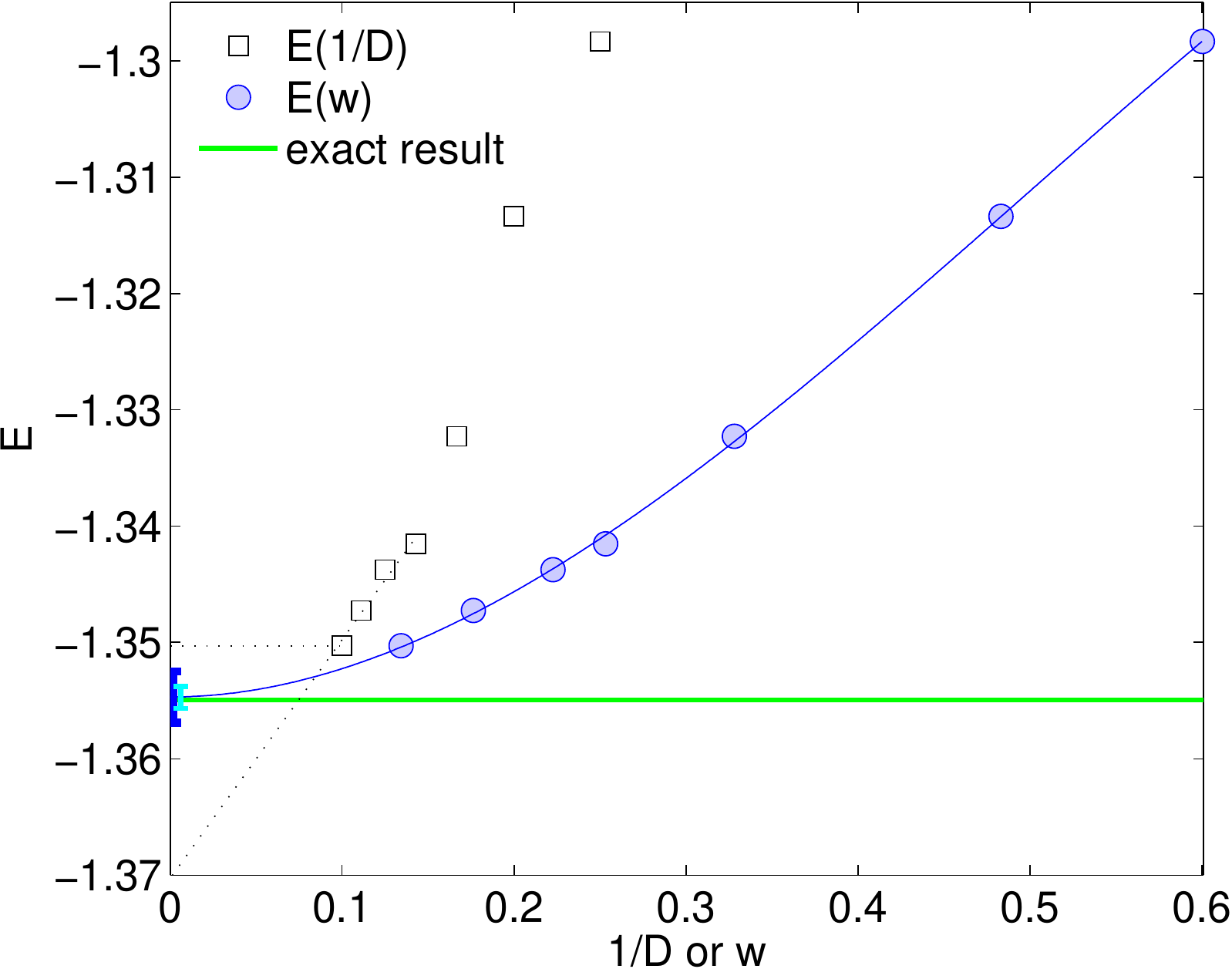} 
\caption{(Color online) iPEPS energy per site of an exactly solvable model of spinful fermions  compared to the exact result.}
\label{fig:pair}
\end{center}
\end{figure}

The iPEPS energies shown in Fig.~\ref{fig:pair} decrease rapidly with increasing $D$. As a consequence, the linear extrapolation in $1/D$ leads to an estimate with a very large error range, $-1.36\pm0.01$. 

In contrast, performing an extrapolation in the truncation error yields a much better estimate, $E_w=-1.3547 \pm 0.002$, or $\tilde E_w = -1.3547 \pm 0.001$ when taking the average over several fits. 

This example  illustrates that estimating the correct energy  based on a $1/D$ extrapolation can be hard, even though the exact value lies not  far from the energy at the largest bond dimension. Thanks to the much smoother behavior of the $E(w)$ curve one can obtain a much more accurate estimate using an extrapolation in $w$.

\bibliographystyle{apsrev4-1}
\bibliography{../bib/refs}

\end{document}